# Chiral Negative-Index Metamaterials in Terahertz

Shuang Zhang[1], Yong-Shik Park[1], Jensen Li[1], Xinchao Lu[2], Weili Zhang[2] and Xiang Zhang[1,3&]

[1] *5130 Etcheverry Hall, Nanoscale Science and Engineering Center, University of California, Berkeley, California 94720-1740, USA*

[2] *School of Electrical and Computer Engineering, Oklahoma State University, Stillwater, Oklahoma 74078, USA*

[3] *Materials Sciences Division, Lawrence Berkeley National Laboratory, 1 Cyclotron Road Berkeley, California 94720*

**Abstract:** Negative index metamaterials (NIMs) give rise to unusual and intriguing properties and phenomena, which may lead to important applications such as superlens, subwavelength cavity and slow light devices[1, 2, 3, 4, 5, 6, 7]. However, the negative refractive index in metamaterials normally requires a stringent condition of simultaneously negative permittivity and negative permeability. A new class of negative index metamaterials - chiral NIMs, have been recently proposed[8, 9, 10]. In contrast to the conventional NIMs, chiral NIMs do not require the above condition, thus presenting a very robust route toward negative refraction. Here we present the first experimental demonstration of a chiral metamaterial exhibiting negative refractive index down to n=-5 at terahertz frequencies, with only a single chiral resonance. The strong chirality present in the structure lifts the degeneracy for the two circularly polarized waves and relieves the double negativity requirement. Chiral NIM are predicted to possess intriguing electromagnetic properties that go beyond the traditional NIMs, such as opposite signs of

---

[&] To whom correspondence should be addressed. E-mail: xiang@berkeley.edu

refractive indices for the two circular polarizations and negative *reflection*[11]. The realization of terahertz chiral NIMs offers new opportunities for investigations of their novel electromagnetic properties, as well as important terahertz device applications.

Optical activity in chiral materials is of broad interest due to its wide applications in biology and chemistry. Chirality introduces cross coupling between the electric and magnetic dipoles, and thus lifts the degeneracy between the two circularly polarized waves by increasing the refractive index for one circular polarization and reducing the index for the other. Consequently, negative refractive index can be obtained in a chiral metamaterial without the stringent requirement of simultaneous negative permittivity ($\varepsilon$) and permeability ($\mu$)[8]. Interestingly, chiral metamaterial can be designed such that the negative refraction works only for one circularly polarized wave, while for the other circular polarization the index is positive. This gives rise to some interesting phenomena that conventional NIMs do not exhibit, such as *negative reflection* for electromagnetic waves incident onto a mirror embedded in such a medium[11]. Furthermore, in the special case where two circularly polarized waves having refractive indices of the same amplitude but opposite signs, the light incident onto the mirror would be reflected back at exactly the same direction[12]. This phenomenon of time reversal is similar to that of light reflected from a phase conjugate mirror, but without involving nonlinearity.

Terahertz is a unique frequency range with many important applications such as security detection and gas phase molecule sensing[13]. However, the devices for manipulating the terahertz wave are considerably limited. Consequently, the development of artificial materials with unusual optical properties at this frequency region is especially important. Recently, the development in metamaterial research has led to the achievement of unusual optical functionalities at terahertz frequencies [14,15,16,17,18,19]. However, due to the complexity of the chiral metamaterial geometry, experimental realizations of chiral NIMs at the terahertz and even higher frequencies still remain major challenges. Although chiral metamaterials were recently studied at the microwave, terahertz and optical frequencies using a simplified bilayer configuration, but no evidence of negative refractive index has been shown in these works [20,21,22,23]. Here we experimentally demonstrate the first negative index chiral metamaterials operating in the terahertz frequencies. This would open doors to exploration of the interesting properties associated with chiral NIMs, as well as broad device applications at terahertz frequencies.

The chiral metamaterial design is based on a vertical metallic chiral resonator, in which the chirality is introduced by tilting the loop of the resonator out of the plane with its gap [Fig. 1(a)]. The chiral resonator is equivalent to a micro-sized inductor-capacitor (LC) resonant circuit, with the inductor formed by the loop and the capacitor formed between the two bottom metal strips [Fig. 1(b)]. Oscillating current flowing through the metal loop can be excited by either an electric field across the gap or a magnetic field perpendicular to the loop, which in turn generate strong electric and magnetic responses[24]. Therefore, this structure can be considered as the combination of an electric dipole and a magnetic dipole, as indicated in Fig. 1(b). Since the electric and magnetic dipoles share the same structural resonance, the excitation of one dipole would inevitably lead to the excitation of the other. Due to the fact that the angle between directions of the two dipoles is small, a strong chiral behavior is expected, which, with properly designed geometric parameters, will lead to negative refraction for circularly polarized waves.

We have fabricated large scale (1.5 cm by 1.5 cm) chiral negative index metamaterials. The SEM images of the structure are shown in Figs. 1(c) and (d). The chiral metamaterials is characterized by terahertz-time domain spectroscopy (THz-TDS)[25, 26]. In the transmission measurement, the chiral sample was placed midway between the transmitter and receiver modules at the waist of terahertz beam [Fig. 2(a)]. Two free standing metal wire polarizers were employed, one in front of and one after the sample to measure the transmission of the same polarization state as that of the incident wave $t_1$ (P1//P2) and that of the perpendicular polarization state $t_2$ (P1⊥P2). The complex coefficients for the transmissions can be obtained by taking the Fourier transform of the time signal, and calibrated over a bare silicon wafer. Owing to the four fold rotational symmetry of the sample, the transmission properties do not rely on the orientation of the sample relative to the polarization of the incident wave, which was confirmed by additional transmission measurements with sample rotated by 90º. For left and right circularly polarized beams the transmission coefficients $t_L$ and $t_R$ can be inferred by,

$$t_L = t_1 - it_2$$
$$t_R = t_1 + it_2 \qquad (1)$$

The measured transmission amplitudes and phases for $t_1$ and $t_2$ are shown in Figs. 2(b) and 2(c). A resonance occurs around 1 THz, exhibiting a dip in $t_1$ and a peak in $t_2$, indicating a strong chiral behavior that leads to the conversion of a large portion of the energy to the other linear polarization. The resonance is accompanied by a steep slope in the relative phases of the transmissions [Fig. 2(c)]. In contrast to the transmission spectra, the reflectance of the chiral metamaterials for terahertz wave does not show pronounced features. The reflectance amplitude is around 0.6 in the range of frequency from 0.2 to 2 THz, and the reflectance phase is close to -$\pi$, which is characteristic of the phase change that light experiences upon reflection from a high dielectric or metal surface. Using equation (1), the transmission for both left circularly polarized beam (LCP) and right circularly polarized beam (RCP) are obtained and shown in Figs. 2(d) and 2(e). The resonance for LCP shows much more pronounced features than that of RCP in the transmission spectra. At resonance, the dip of LCP transmission amplitude approaches almost zero (<6%, or transmission less than 0.4%), which indicates that a beam with a linear polarization transmitting through the sample would be largely converted into a RCP. Therefore, this chiral metamaterial can be used as an effective circular polarization filter at the resonance frequency. The transmission phase of LCP exhibits a much larger phase modulation than that of RCP, which indicates that the refractive index for LCP changes dramatically across the resonance frequency. A full wave simulation using a time-domain simulation software (CST Microwave Studio$^{TM}$) was also carried out to calculate the transmission and reflectance, as shown in Fig. 3. Apparently, a reasonable agreement between the simulation and measurement has been obtained.

The electromagnetic response of the chiral metamaterials can be written as the following constituent equations,

$$\begin{pmatrix} D_x \\ D_y \end{pmatrix} = \varepsilon_0 \varepsilon \begin{pmatrix} E_x \\ E_y \end{pmatrix} + \frac{i}{c_0} \begin{pmatrix} \xi & \xi_{12} \\ -\xi_{12} & \xi \end{pmatrix} \begin{pmatrix} H_x \\ H_y \end{pmatrix}$$

$$\begin{pmatrix} B_x \\ B_y \end{pmatrix} = \frac{i}{c_0} \begin{pmatrix} -\xi & \xi_{12} \\ -\xi_{12} & -\xi \end{pmatrix} \begin{pmatrix} E_x \\ E_y \end{pmatrix} + \mu_0 \mu \begin{pmatrix} H_x \\ H_y \end{pmatrix} \quad (2)$$

where ε and μ are the effective permittivity and permeability, $\xi$ and $\xi_{12}$ describe the excitation of electric (magnetic) dipoles by the magnetic (electric) field along the same and the perpendicular directions, respectively. The relation $\xi_{12} = \xi \tan\theta$ holds due to the fact that the magnetic dipole and electric dipole form an angle θ in the chiral resonator [Fig. 1(b)]. The chiral parameter ξ lifts the degeneracy and leads to different refractive indices $n_{L/R}$ for the two circularly polarized waves, while the bi-anisotropic parameter $\xi_{12}$ results in different effective impedances $Z_+$ and $Z_-$ of the metamaterial for light along +z (light incident from air) and -z (light incident from substrate), respectively [27]. $\varepsilon$, $\mu$, $\xi$ and $\xi_{12}$ are related to $n_{L/R}$ and $Z_{+/-}$ by

$$(n_{L/R} \mp \xi)^2 = \varepsilon\mu - \xi_{12}^2$$
$$Z_\pm = \frac{\mu}{\sqrt{\varepsilon\mu - \xi_{12}^2} \pm i\xi_{12}} \quad (3)$$

The effective parameters of the chiral metamaterial has been extracted by solving the Fresnel's equation using the *measured* transmission and reflectance spectra, and taking into account the substrate. The effective indices for both LCP and RCP are shown in Fig. 4(a) and 4(b). The real part of refractive index for the left handed wave shows a strong modulation at the resonance frequency, and reaches negative values over the frequency range between 1.06 to 1.27 THz, with a minimum index below -5. This, to the best of our knowledge, is the first experimental demonstration of a negative index chiral metamaterial. On the other hand, the refractive index for RCP shows a much smaller modulation at the resonance frequency and remains positive over the whole frequency range. This is due to the cancellation of the resonant feature of refractive index between the contribution from permittivity and permeability, and that from the chirality, as indicated by Eq. (3). Far from the resonance, the difference between the refractive indices of RCP and LCP tends to diminish, showing the significant role that the structural LC resonance plays in achieving the strong chiral behavior. The effective permittivity and permeability are extracted and shown in Figs. 4(c) and 4(d). Both of them exhibit a Lorentzian lineshape. While the permittivity reaches negative values (with a minimum of -13.7) at resonance, the magnetic resonance is not strong enough to achieve negative permeability. Nevertheless, the existence of strong chirality [Fig. 4(e)] in the metamaterial relieves the condition of simultaneous negative permittivity and permeability, as required by conventional negative index metamaterials.

We have demonstrated the first terahertz chiral artificial materials exhibiting opposite signs for the effective refractive indices of the two circularly polarized waves around the resonance frequency of 1 terahertz. The optical properties of the chiral metamaterials are obtained solely from the experimental measurement. The realization the chiral NIMs offers us the opportunity to explore their unique electromagnetic properties, such as negative refraction and negative reflection. The terahertz metamaterials, when combining with the optical or electrical tuning techniques [16, 17, 18, 19], may lead to terahertz devices which can control the polarization of terahertz wave dynamically.

**Figure captions:**

**Fig. 1 Structure geometry and the equivalent circuit.** (**a**) The schematic of the chiral structure, with some of the dimensions indicated in the figure: L =20μm, h=4.5μm, r=1.6μm, w=4.4μm, g=2.3μm. The thicknesses of the bottom gold strips and the top gold bridge are 0.6 and 0.3 μm, respectively. The bottom gold strips make an angle $\theta = 29.25°$ with the top gold bridge. (**b**) The inductor-capacitor circuit model of the chiral structure, which functions effectively as an electric dipole (purple arrow, along the direction of capacitor) and a magnetic dipole (blue arrow, along the direction of inductor) forming an angle θ. (**c, d**) the SEM images of the chiral metamaterials at tilted angle. The size of the unit cell is 40 μm by 40 μm. The scale bars are 20 μm.

**Fig. 2 The time-domain terahertz measurements of the chiral metamaterials.** (**a**) two-polarizer setup to measure the transmission coefficients. Terahertz wave propagates along +z direction. The polarization of terahertz source and detection are in the horizontal direction (x-axis). The polarizers are tilted 45° with respect to the vertical direction (y-axis). Polarizer 2 is either parallel or perpendicular to polarizer 1, for the measurement of $t_1$ or $t_2$, respectively. (**b**) The amplitudes of $t_1$ and $t_2$ and reflectance r. (**c**) The phases of $t_1$, $t_2$ and r. (**d**) The transmission amplitudes for left handed and right handed circular polarizations. (**e**) The transmission phases for left handed and right handed circularly polarizations.

**Fig. 3**. **The simulated transmission and reflectance of the chiral metamaterials**. In the simulation, the gold is described by the Drude model with plasma frequency $\omega_p = 1.37 \times 10^{16} s^{-1}$ and scattering frequency $\gamma = 2.04 \times 10^{14} s^{-1}$. (**a**) The amplitudes of transmission $t_1$ and $t_2$ for the polarizer P2 parallel to P1 (black) and P2 perpendicular to P1 (red), as well as the amplitude of reflectance (blue) (**b**) The phases of $t_1$ (black), $t_2$ (red) and reflectance r (blue). (**c**) The transmission amplitudes for left handed (black) and right handed circularly polarized waves (red). (**d**) The transmission phases for left handed (black) and right handed circularly polarized waves (red).

**Fig. 4 Experimentally retrieved effective parameters**. (**a**, **b**) The experimentally retrieved real (black) and imaginary (red) parts of the refractive index for left handed and right handed circularly polarized wave, respectively. The shaded area indicates the negative-index region. (**c-e**) the real (black) and imaginary (red) parts of the permittivity ε, permeability μ and the chiral parameter ξ.

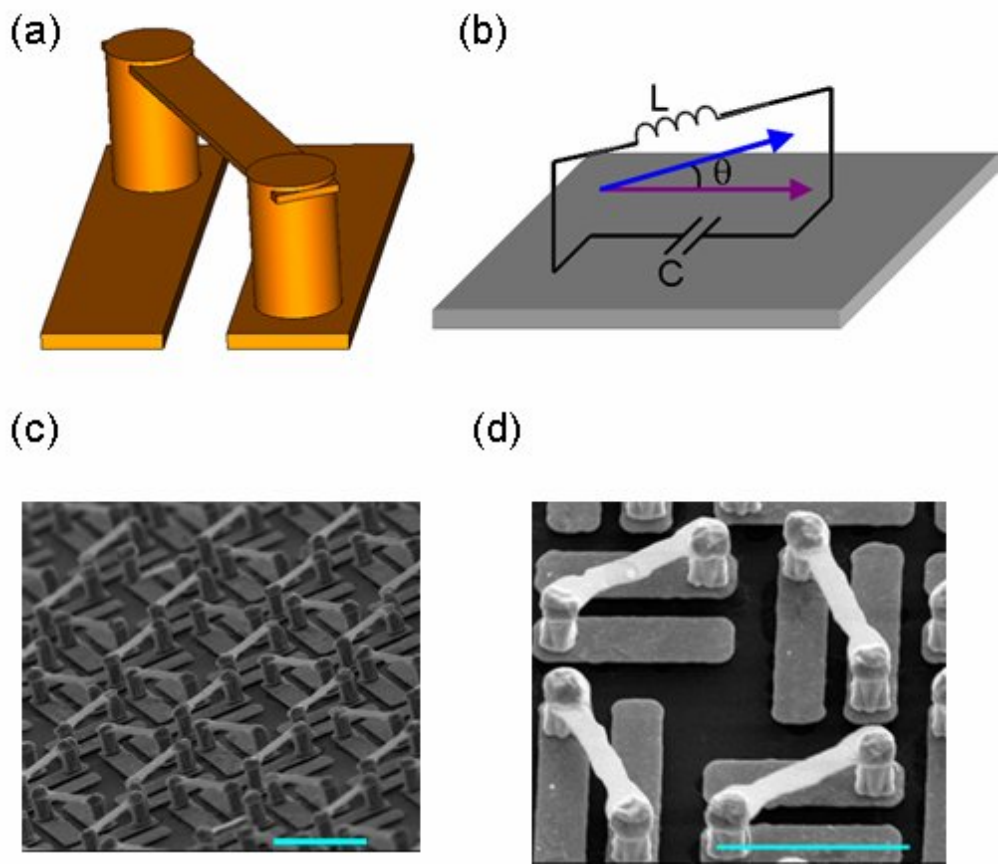

**Fig. 1**

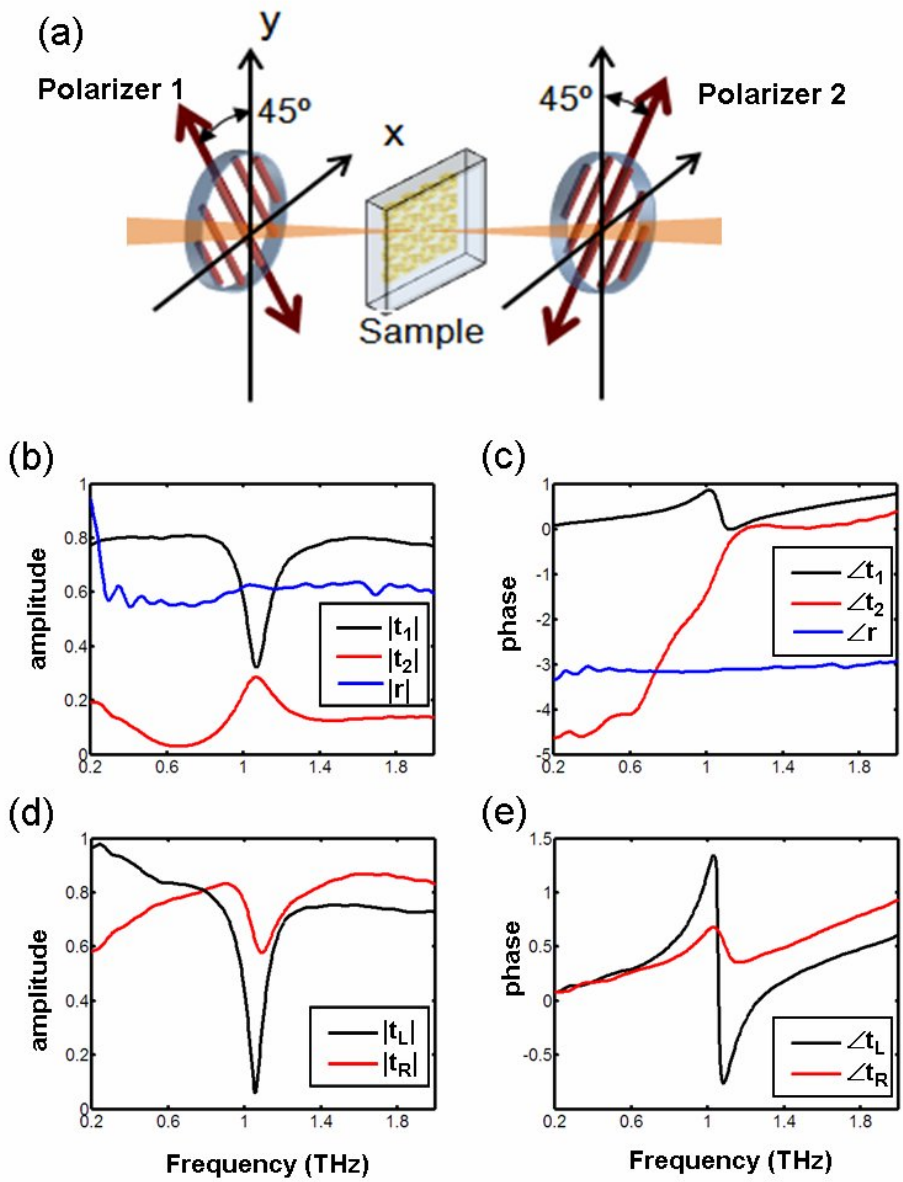

**Fig. 2**

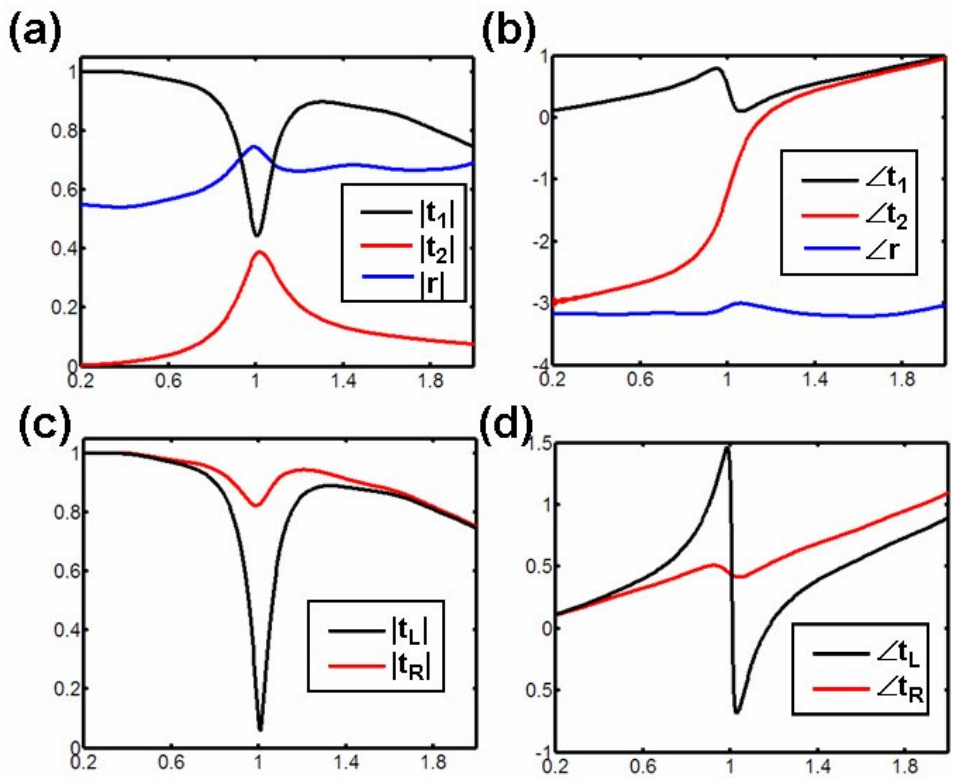

**Fig. 3**

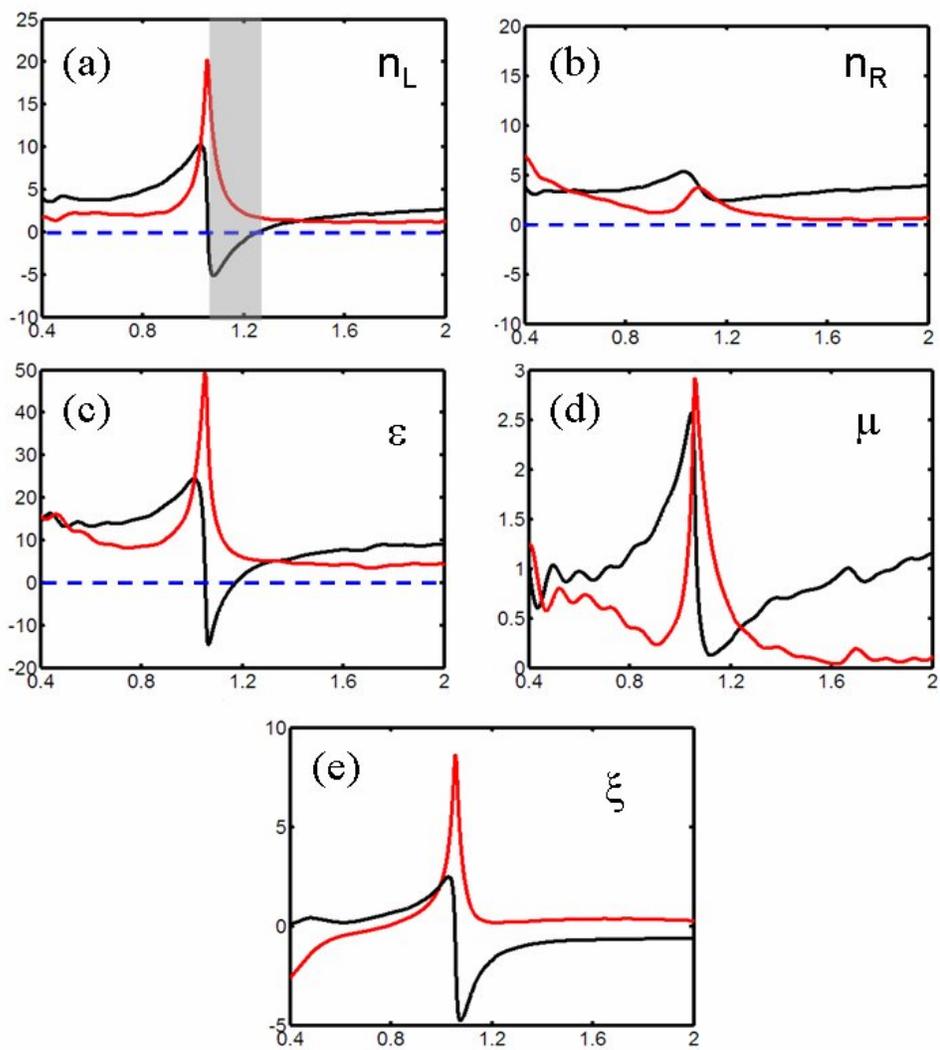

**Fig. 4**